\documentclass[graybox]{svmult}

\usepackage{mathptmx}       % selects Times Roman as basic font
\usepackage{helvet}         % selects Helvetica as sans-serif font
\usepackage{courier}        % selects Courier as typewriter font
\usepackage{type1cm}        % activate if the above 3 fonts are
\usepackage{makeidx}         % allows index generation
\usepackage{graphicx}        % standard LaTeX graphics tool
\usepackage{multicol}        % used for the two-column index
\usepackage[bottom]{footmisc}% places footnotes at page bottom
\usepackage{subfigure}
\usepackage{multirow}
\usepackage[table,xcdraw]{xcolor}
\usepackage[misc]{ifsym}

\makeindex             % used for the subject index
\begin{document}

\title*{Crossing Behaviour of Social Groups: \\Insights from Observations at Non-signalized Intersection}
\titlerunning{Crossing Behaviour of Social Groups} 
\author{Andrea Gorrini, Luca Crociani, Giuseppe Vizzari, Stefania Bandini}
\authorrunning{Gorrini, A., Crociani, L., Vizzari, G., Bandini, S.} 
\institute{Andrea Gorrini (\Letter) \and Luca Crociani \and Giuseppe Vizzari \and Stefania Bandini \at CSAI research center, University of Milano-Bicocca, Viale Sarca 336 - U14, 20126 Milan (Italy).\\\email{{name.surname}@disco.unimib.it} 
\and Stefania Bandini \at RCAST, The University of Tokyo, 4-6-1 Komaba, Meguro-ku, Tokyo 153-8904 (Japan).}
%
% Use the package "url.sty" to avoid
% problems with special characters
% used in your e-mail or web address
%
\maketitle

\abstract*{}

\abstract{Environmental, demographical and psychological factors have a demonstrated impact on risky crossing behaviour. In this work we focus on the potential influence of social factors on the considered phenomenon (i.e. \emph{group crossing decision}). We present the results of a video-recorded observation about the crossing behaviour of singles and dyads at non-signalized intersections. Results showed that crossing behaviour is characterized by three distinct phases: (i) approaching, (ii) appraising (decision making) and (iii) crossing. Dyads walk slower than single pedestrians in all phases. The crossing behaviour of dyads is characterized by the emergence of a leader who takes the decision to cross first, followed by the companion. However, there is no difference between the accepted safety gap of singles and dyads. Understanding factors influencing the crossing decision of social groups represents an important result supporting the development of agent-based simulations of pedestrian-vehicle interactions.}

\section{Introduction}
\label{sec:1}

As highlighted by the WHO \cite{world2015global}, road accidents represent the eighth leading cause of death in the world population: 1.2 million people are killed on roads every year. Despite recent efforts, the measures currently in place to reduce the phenomenon are mainly aimed at protecting car occupants. However,   pedestrians are some of the most vulnerable road users, with a percentage of fatalities corresponding to 22\% of the overall victims (26\% in EU, 14\% in USA).

To effectively contrast the social costs of pedestrian-car accidents it is necessary to identify the determining factors of risky crossing decisions by means of a multi-disciplinary approach (e.g., traffic psychology, transportation engineering, safety science, computer science). This is aimed at supporting public institutions in the design of effective and safe infrastructures and traffic management solutions.

In this context, the results presented in \cite{sisiopiku2003pedestrian,perumal2014study} highlighted for instance the impact of the physical features of the infrastructure on crossing behaviour (e.g., road width, traffic volumes, type of intersection). Other researches focused on the complex coordination of the motor and cognitive abilities involved in crossing behaviour, such as:  locomotion \cite{sisiopiku2003pedestrian}; perception and attention \cite{hamed2001analysis}; attitude towards hazardous situations \cite{evans1998understanding}. Other studies \cite{dommes2015towards,PED2016crossing} showed the impact of ageing on crossing behaviour, due to the progressive decline of these functions.

These works showed the relevant impact of environmental, demographical and psychological factors on risky crossing behaviour. However, there is still a lack of knowledge about the impact of social factors on the phenomenon, due to limited or controversial results. Urban cross-walks are characterized indeed by the presence of multiple pedestrians, and people often cross the street in the presence of other familiar or unfamiliar pedestrians. 

The contributions about this topic can be classified according two different approaches. On one hand, the results presented in \cite{harrell1991factors,ren2011crossing} showed that the presence of other pedestrians is associated with a reduction of cautiousness in crossing decision at signalized cross-walks. Moreover, the results presented in \cite{faria2010collective} showed that pedestrians imitate the behaviour of others to judge the gap from oncoming vehicles, with potentially dangerous crossing decisions due to an overestimation of crossing possibilities. This phenomenon is based on the power of group to cause a diffusion of responsibility \cite{festinger1952some}, as each member feels that the responsibility for violating the norms and for taking risky crossing decisions is shared with the rest of the group.  

On the other hand, the results collected by \cite{rosenbloom2009crossing,Cavallo2017} showed that people are more lax in risky crossing decisions standing by their own than when standing in groups. These findings can be explained considering the theory of social control \cite{hirschi1994generality}, which describes the mechanism behind obedient behaviour as the motivation to be rewarded just for being conformist to the group norms. 

Starting from these assumptions, the present work was focused on studying the crossing behaviour of social groups (groups composed of familiar pedestrians, such as relatives and friends) through a video-recorded observation at a non-signalized intersection. The research was aimed at comparing data about the crossing behaviour of single pedestrians and two-members groups (i.e. \emph{dyads}): the simplest and most frequently observed type of group. 

The methodology which sets the current work is presented in Section \ref{sec:2}, with reference to data collection and data analysis. The results of the observation (e.g., traffic volumes, Level of Service, crossing phases, speeds, safety gap) are presented in Section \ref{sec:3}. The paper concludes with remarks about the achieved results and their future use for the further development of a computer-based simulator of pedestrian/vehicles interaction in urban contexts~\cite{FelicianiIA2017}.

\section{Data Collection}
\label{sec:2}

\begin{figure}[t!]
\begin{center}
\includegraphics[width=\textwidth]{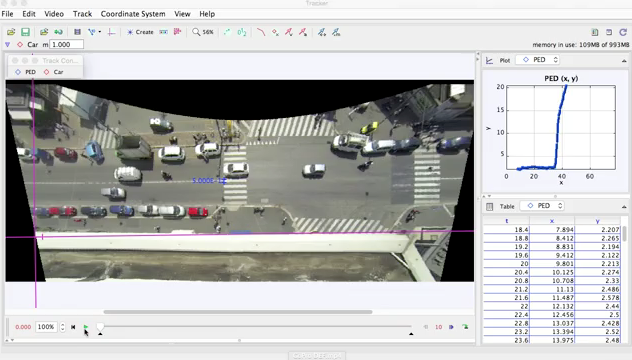}
\caption{A screen shot from the tracker tool used for data analysis.}
\label{fig:scenario}
\end{center}
\end{figure}

\begin{table}[t!]
\small
\begin{tabular}{p{4cm}p{7.5cm}} 
\hline\noalign{\smallskip}
\emph{Locomotion behaviour} & - high spatial cohesion and coordination while walking \\
& - waiting dynamics to regroup in case of separation \\
& - leader/follower dynamics in sudden changes of direction \\ 
\hline\noalign{\smallskip}
\emph{Verbal behaviour}        &  - talking while walking \\ 
\hline\noalign{\smallskip}
\emph{Non Verbal behaviour}     &  - physical contact \\   
& - body and gaze orientation to the each other \\   
& - gesticulation while talking and/or indicating \\ 
\hline\noalign{\smallskip}
\end{tabular}
\caption{The check-list used by the coders for identifying social groups.}
\label{tab:checklist} 
\end{table}

The video-recorded observation~\cite{PED2016crossing} was performed in 2015 (from 10:45 am, 73 minutes in total) at a non-signalized intersection in Milan (Italy), which was selected by means of a preliminary analysis related to the localisation of road traffic accidents. Results showed that the area was characterised by a high number of pedestrian/car accidents in the past years. 

A first phase of manual counting activity allowed to estimate and characterise the observed traffic volumes. An \emph{ad hoc} designed check-list (see Tab. \ref{tab:checklist}), comprising a set of locomotion and communication indicators, was used to support the annotators in the identification of social groups from the video-images. Then, a time stamping activity allowed to measure the Level of Service in the observed intersection. 

A second phase of data analysis was based on the use of the open source software Tracker Video Analysis and Modelling Tool (see Fig. \ref{fig:scenario}). The tracker allowed to correct the distortion of the video images (due to due to the wide lens and the nearly zenith perspective of the camera), and to track a sample of pedestrians considering one frame every ten (every 0.4 sec). The data set (including the \emph{X}, \emph{Y} coordinates and the associated frames \emph{t}) was exported for data analysis. 

\section{Results}
\label{sec:3}

\subsection{Traffic Volumes}

The observed vehicular traffic volumes (1379 veh, 18.89 veh/min) were constituted for the majority by cars (67\%). The direction of movements of vehicles was equally distributed. The majority of pedestrians (585 ped, 8.01 ped/min) were singles (65\%). Dyads were a significant portion of the total counted pedestrians: 26\% dyads; 8\% triples; 1\% other. A Cohen\rq s Kappa analysis showed a moderate agreement between the two independent coders in the identification of social groups (K = 0.47).

\subsection{Level of Service}

The Level of Service (LOS) \cite{HCM2010} describe the degree of efficiency of an intersection, by measuring the additional travel time experienced by drivers and pedestrians as they travel/walk through a road segment. At non-signalized intersections LOS E is the minimum acceptable design standard. 

The LOS have been estimated by time stamping the delay of drivers due to vehicular traffic and crossing pedestrian flows (time for deceleration, queue, stopped delay, acceleration), and the delay of pedestrians due to drivers\rq\ non compliance to their right of way on zebra (waiting, start-up delay). Results showed that the average delay of drivers (3.20 s/veh $\pm$ 2.73 sd) and pedestrians (1.29 s/ped $\pm$ .21 sd) corresponded to LOS A: nearly all drivers found freedom of operation; low risk-taking in crossing behaviour. 

\subsection{Speeds and Crossing Phases}

A sample of 49 adult pedestrians (27 singles and 11 dyads, from about 18 y.o. until 65 y.o.) was selected for video-tracking analysis. The sample was selected avoiding situations such as: platooning of vehicles on the roadway, the joining of pedestrians already crossing, and in general situations influencing the direct interaction between pedestrians and drivers. Part of the selected crossing episodes was characterised by the multiple interaction between the crossing pedestrian and two cars oncoming from the near and the far lane.
Pedestrians\rq\ gender was balanced, but not considered for data analysis.  

The speeds  pedestrians have been analysed among the time series of video frames. The trend of speeds was analysed by calculating the difference between the moving average (MA, time period length: 0.8 s) and the cumulative average (CA) of the entire frames series. This allowed to smooth out short-term fluctuations of data and to highlight longer-term deceleration/acceleration trends. According to results, crossing behaviour is defined as composed of three distinctive phases:

\begin{enumerate}
\item \emph{approaching phase}: the pedestrian travels on the side-walk with a stable speed (Speed MA - CA $\simeq$ 0);
\item \emph{appraising phase}: the pedestrian approaching the cross-walk decelerates to evaluate the distance and speed of oncoming vehicles (safety gap). We decided to consider that this phase starts with the first value of a long-term deceleration trend (Speed MA - CA $<$ 0);
\item \emph{crossing phase}: the pedestrian decides to cross and speed up. The crossing phase starts from the frame after the one with the lowest value of speed before a long-term acceleration trend (Speed Min).
\end{enumerate}

\begin{figure}[t!]
\begin{center}
\subfigure{\includegraphics[width=0.9\textwidth]{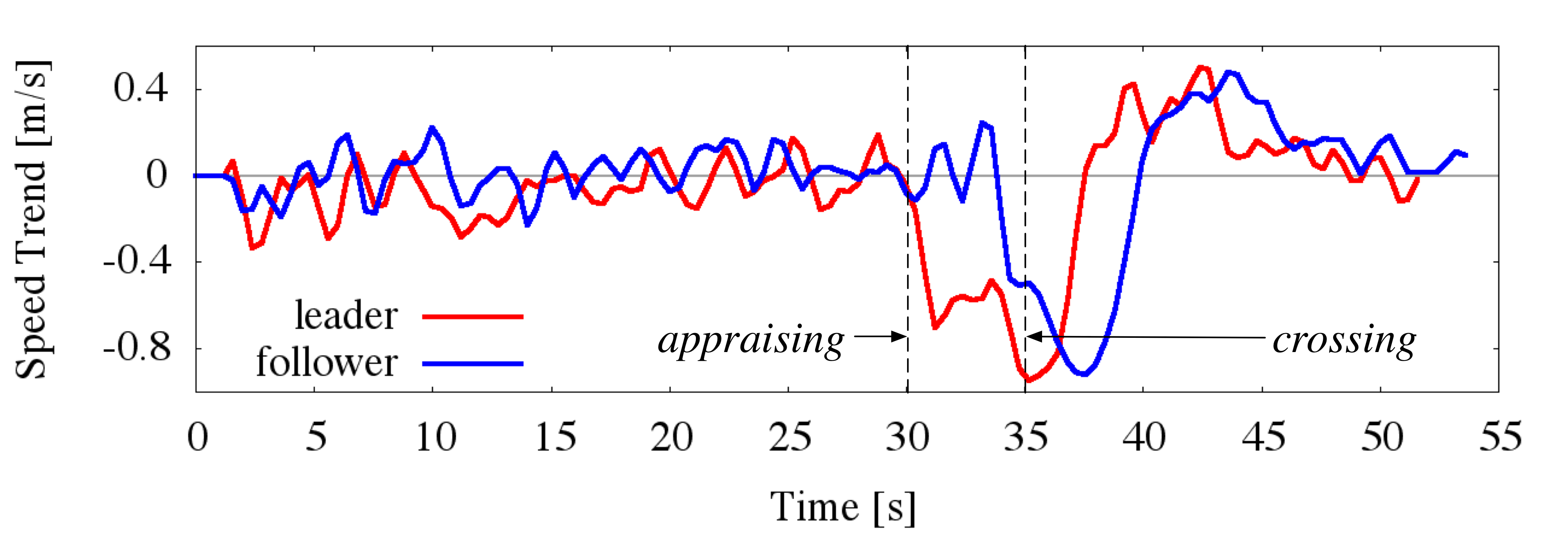}}
\caption{An exemplification of the trend analysis performed on the time series of dyads\rq\ speeds.}
\label{fig:phases}
\end{center}
\end{figure}

\begin{figure}[t]
\begin{center}
\subfigure[Singles]{\includegraphics[width=.49\textwidth]{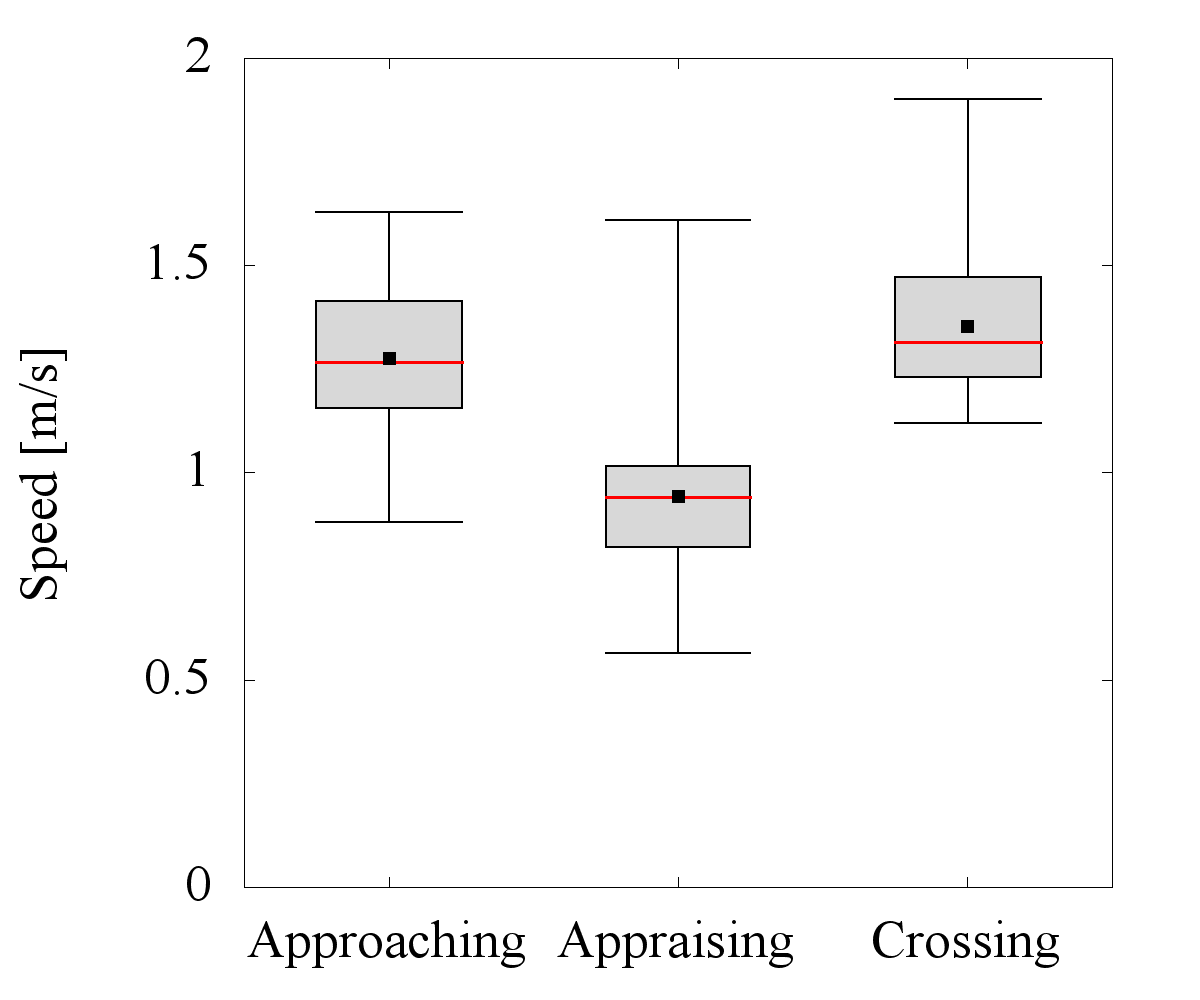}}
\subfigure[Dyads]{\includegraphics[width=.49\textwidth]{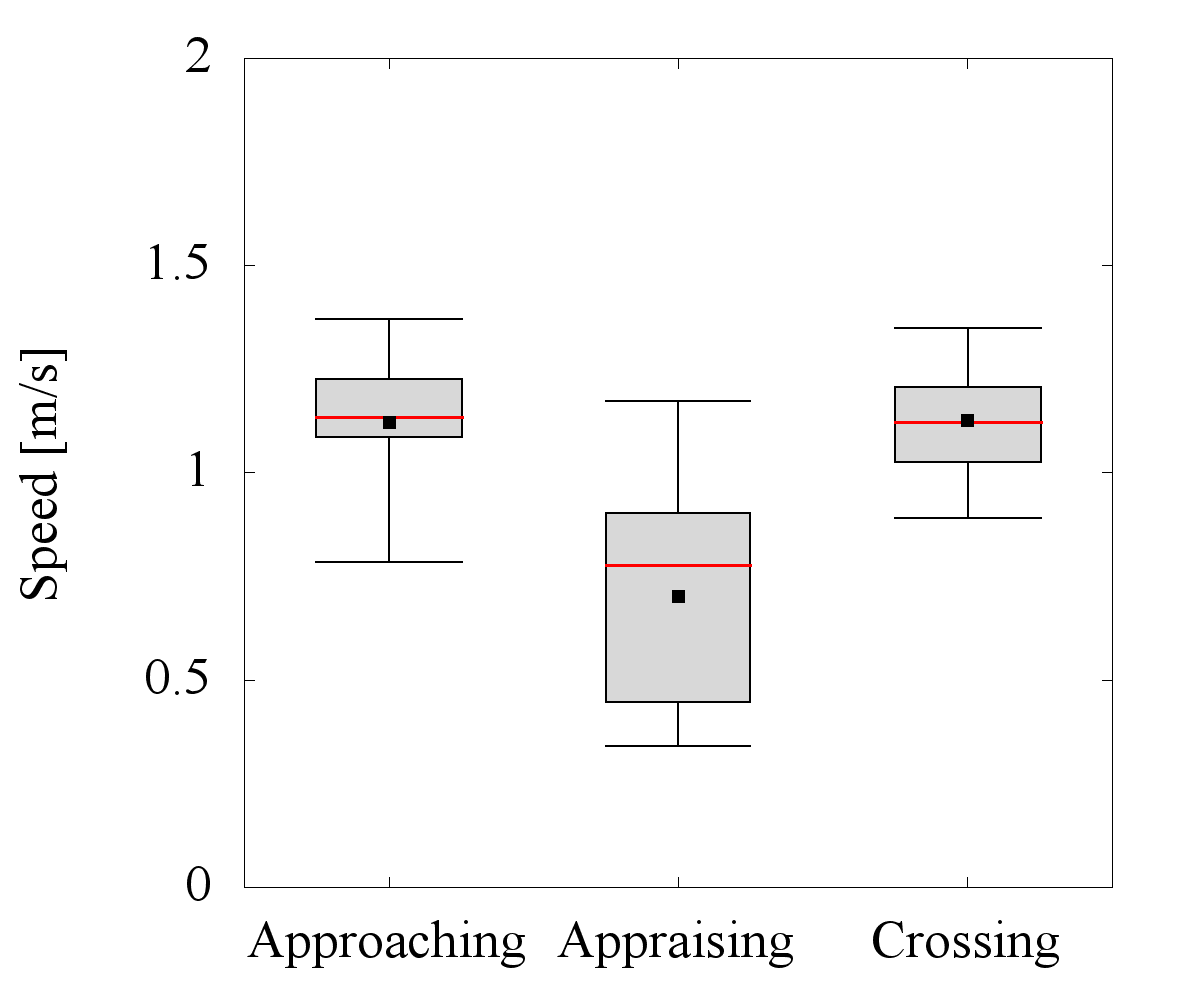}}
\caption{The speed of single pedestrians and dyads among the crossing phases.}
\label{fig:candle_speed}
\end{center}
\end{figure}

A two-way ANOVA\footnote{All statistics have been executed by using \emph{IBM SPSS Statistics 24}, at the p $<$ 0.01 level.} showed a significant difference among the speeds of pedestrians while approaching, appraising and crossing [F(2,146) = 66.981, p $<$ 0.001], and a significant difference between the speeds of singles and dyads [F(1,146) = 40.652, p $<$ 0.001] (see Tab. \ref{tab:speed} and Fig. \ref{fig:candle_speed}). A series of post hoc Tukey test showed a non significant difference between the speeds of all pedestrians while approaching and crossing (p $>$ 0.05). The difference between the speeds of singles and dyads was significant among all the crossing phases (p~$<$~0.001). 

Results showed that the crossing behaviour of dyads (see Figure \ref{fig:phases}) is characterised by the emergence of a leader who completes the appraising first and takes the decision to cross, followed by the companion. An independent samples t-test analysis showed that the difference between the speeds of group leaders and followers was not significant among all the three crossing phases (p $>$ 0.05).

Results demonstrated that the crossing behaviour of both singles and dyads is based on a significant deceleration in proximity of the curb to evaluate the distance and speed of oncoming vehicles to safely cross. Dyads walked in average 17\% slower than singles among the three crossing phases, decelerating the 11\% more than singles while appraising. 

\begin{table}[t!]
\centering
\small
\caption{The speed of singles and dyads among the crossing phases.}\vspace{0.5cm}
\label{tab:speed}
\begin{tabular}{p{3.5cm}p{4cm}p{4cm}}
\hline\noalign{\smallskip}
\textbf{Crossing Phases}      & \textbf{Speed of Singles}     & \textbf{Speed of Dyads}   \\ 
\hline\noalign{\smallskip}
Approaching speed & 1.277 m/s $\pm$ 0.182 sd & 1.123 m/s $\pm$ 0.155 sd \\ 
\hline\noalign{\smallskip}
Appraising speed & 0.943 m/s $\pm$ 0.214 sd  & 0.702 m/s $\pm$ 0.280 sd  \\ 
\hline\noalign{\smallskip}
Crossing speed    & 1.354 m/s $\pm$ 0.181 sd & 1.127 m/s $\pm$ 0.131 sd \\ 
\hline\noalign{\smallskip}
\end{tabular}
\end{table}

\subsection{Accepted Safety Gap}

The term \emph{safety gap} denotes the ratio between the pedestrians\rq s evaluation of the distance of an approaching vehicle and its speed (not taking into account acceleration/deceleration trends) to decide to safely cross avoiding collision. 

A one-way ANOVA showed that the safety gap accepted by singles (3.982 s $\pm$ 2.549 sd), group leaders (4.355 s $\pm$ 2.491 sd) and followers (4.242 s $\pm$ 2.585 sd) was not significantly different (p $>$ 0.05) (see Table \ref{tab:gap}). Moreover, a one-way analysis of variance showed that the time duration of the appraising phase of singles (2.785 s $\pm$ 1.474 sd), group leaders (3.527 s $\pm$ 2.297 sd) and followers (3.164 s $\pm$ 1.682) were not significantly different (p $>$ 0.05). 

Results showed also that, compared to leaders, followers started appraising with a delay of 1.055 s ($\pm$ 1.032 sd) and they decide to cross with a delay of 0.691 s ($\pm$ 0.671 sd). Further analysis about the relative positions of group members showed that the large majority of group leaders preceded followers while appraising, and that all leaders preceded followers when they decided to cross. Results highlighted that the leaders were able to better evaluate the distance and speed of oncoming vehicles according to the position, and so they take the decision to cross first followed by the companion. 

\begin{table}[t!]
\centering
\small
\caption{The safety gap accepted by singles, group leaders and followers.}
\label{tab:gap}
\begin{tabular}{p{2.2cm}p{3cm}p{3cm}p{3cm}}
\hline\noalign{\smallskip}
\textbf{}                 & \textbf{Singles}     & \textbf{Group Leaders}  & \textbf{Followers}     \\
\hline\noalign{\smallskip}
Safety Gap       & 3.982 s $\pm$ 2.549 sd     & 4.355 s $\pm$ 2.491 sd & 4.242 s $\pm$ 2.585 sd \\
\hline\noalign{\smallskip}                 
Vehicle Distance & 16.399 m $\pm$ 7.753 sd    & 19.152 m $\pm$ 9.967 sd & 17.325 m $\pm$ 9.583 sd \\
\hline\noalign{\smallskip}
Vehicle Speed    & 16.534 km/h $\pm$ 6.737 sd & 16.525 km/h $\pm$ 5.662 sd & 15.695 km/h $\pm$ 4.479 sd \\
\hline\noalign{\smallskip}  
Appraising Time    & 2.785 s $\pm$ 1.474 sd & 3.527 s $\pm$ 2.297 sd & 3.164 s $\pm$ 1.682 \\
\hline\noalign{\smallskip}   
\end{tabular}
\end{table}

\section{Final Remarks}
\label{sec:4}

A video-recorded observation was performed at a non-signalized intersection to study the crossing behaviour of social groups, comparing data among single pedestrians and dyads. Trend analyses on the speeds of the tracked pedestrians showed that crossing behaviour is characterised by three distinct phase: approaching, appraising and crossing. In particular, the appraising phase consists on a significant deceleration in proximity of the curb to evaluate the distance and speed of oncoming vehicle to safely cross (i.e. \emph{decision making}). 

Results showed that dyads walked in average 17\% slower than singles among the three crossing phases, decelerating the 11\% more than singles while appraising. Group crossing behaviour is based on the emergence of a leader who completes the appraising first and decide to cross, followed by the companion. However, followers did not merely imitate the crossing behaviour of the leaders in judging the safety gap from oncoming vehicles, since no significant difference was found in the time duration of the appraising phase between them. 
Finally, no significant difference was found in the safety gap accepted by singles and dyads.

Future works will be aimed at enlarging the sample of tracked pedestrians for data analysis, including also groups composed by unfamiliar pedestrians. Moreover, we consider the possibility to test the impact of social imitation among unfamiliar pedestrians on risky crossing behaviour, such as non-compliant jaywalking behaviour out of zebra-striped crossing.

While no significant difference has emerged in comparing the crossing decision of singles and dyads, the presented results could be of notable interest for those involved in modelling and simulation of urban interactions. In particular, the results of the observation will be used to support the further development of a computer-based simulator of pedestrian/vehicles interactions at non-signalised intersections~\cite{FelicianiIA2017}, focusing on modelling heterogeneous crossing speed among pedestrians.

\begin{acknowledgement}
The Italian policy was complied in order to exceed the ethical issues about the privacy of the people recorded without their consent. The authors thank Fatima Zahra Anouar for her fruitful contribution in data analysis. The recorded video tape and the annotated data set are available for research purposes.
\end{acknowledgement}

%The reference style for working with bibtex is 
%\bibliographystyle{spmpsci}
%then input your bib file
%\bibliography{tgfbibs}

%alternatively, you can input the references manually. Keep to the format described in 
%Mathematical and Physical Sciences.pdf
%\input{referenc}

\end{document}